\documentclass{article}
\usepackage{array, latexsym, graphicx,times,amsfonts,picture,color,multirow,epsf,epsfig,mathtools}
\usepackage{amsmath}
\usepackage{verbatim}
\usepackage{blkarray}
\usepackage{setspace}
\usepackage{subfigure}
\usepackage{breqn}
\usepackage{listings}
\usepackage{enumitem}

\usepackage[T1]{fontenc}
\usepackage{textcomp}

\usepackage{hyperref}
\begin{document}
\title{Another Look at ALGORAND}
\author{Yongge Wang\\ UNC Charlotte}
\maketitle

\begin{abstract}
ALGORAND is a celebrated public ledger technology designed by Dr. Micali and his collaborators. By the end of year 2018, it has
raised US\$64 million in funding from venture capital firms. In this paper, we identify several design 
flaws of the ALGORAND protocol.  In particular, we show that the claimed (proved) fork-free property is not true and 
several assumptions in ALGORAND are not realistic in practice. 
The ALGORAND wiki page \url{https://golden.com/wiki/Algorand} claims that 
``{\em the probability of a fork in the protocol is estimated at 
1/1,000,000,000 and therefore blocks can be considered final upon validation}''. 
However, our first attack in this paper shows that 
a malicious adversary who controls less than 1/3 of the users (or money units) 
could fork the ALGORAND chain very easily. Our second attack shows that a malicious adversary
could use a bribery attack to fork  the ALGORAND chain very easily also. 
Furthermore, we show that the celebrated
Byzantine Agreement component in ALGORAND is not necessary. The Byzantine Agreement is the most 
expensive part and one of the most  innovative parts in the ALGORAND protocol. It is used to avoid
forks in ALGORAND. We show that a simple majority vote could be used to achieve the same property that Byzantine
Agreement achieves in ALGORAND under the same network assumption.
\end{abstract}

\section{Introduction}
In a digital society, it would be convenient to have a digital payment system or to have a digital currency system. It is generally easy to design an electronic cash system using public key infrastructure (PKI) systems. But PKI-based electronic cash is also easy to trace. Theoretically, banknotes could be traced using sequence numbers, though there is no convenient infrastructure to trace banknote sequence numbers back to users. Banknotes thus maintain sufficient anonymity. 

Several researchers have designed anonymous electronic cash systems. The early effort includes
Chaum's online untraceable payment system \cite{chaum1983blind} based on Chaum's blind signatures 
and Chaum, Fiat, and Naor's \cite{chaum1990untraceable} electronic cash
that does not need the bank to be online. However, these systems have not attracted enough interest from
the society and they have never been adopted. The situation has changed since the cryptographic 
currency Bitcoin was introduced in the paper \cite{nakamoto2008bitcoin} by
a pseudonym ``Satoshi Nakamoto''. Since 2009, the implementation of Bitcoin 
has been in operation and it has been widely adopted as one of the major 
cryptographic currency on the market now. Bitcoin used Forth-like Scripts 
for writing smart contracts. In order to increase the smart contract capability, 
Ethereum used Turing-complete programming language {\em Solidity} for 
its smart contract design.

In the Bitcoin system,  one can achieve system consensus under the assumption 
that more than 51\% computational power is honest. This ``contradicts''
the classical results in Byzantine Agreement which requires at least 2/3 of the participants to be honest for achieving consensus.
However, Bitcoin has several inherent technical challenges. First, Bitcoin uses  proof of work
to generate new blocks.  This requires a lot of computation and wastes a lot of energy.
Secondly, due to the enormous amount of computational power and energy requirements,
it is not profitable for regular desktop computers to mine new Bitcoin blocks.
Thus the major computing powers for Bitcoin block generation are currently from a few mining pools
(in particular, Chinese mining pools control more than 75\% of the Bitcoin network's collective hashrate in 2019)
and the assumption of honest majority computing power may no longer be valid.
Thirdly, Bitcoin block chain may fork once a while. Thus one needs to wait for a few blocks to make sure that 
her payment becomes permanent on the block chain.

The ingenious design of Bitcoin has inspired a lot of fruitful research. Several researchers
have introduced proof-of-stake or proof-of-``anything'' block chain techniques
to address the challenges faced by Bitcoin. One of these celebrated results is ALGORAND 
by Micali and his collaborators (see, e.g., \cite{chen2016algorand,gilad2017algorand}).

ALGORAND works both in permissionless and permissioned environments. For  the permissioned environments,
it assumes at at least $2/3$ of the users are honest and for the permissionless environments, it assumes that at least $2/3$ of 
the money belongs to honest users. Blocks in ALGORAND could be generated in less than
40 seconds in experiments (see, e.g., \cite{chen2016algorand}). The major techniques employed by ALGORAND include:
\begin{itemize}
\item A fast constant round Byzantine Agreement (BA) protocol
\item A secret cryptographic sortition process to select a subset of secret users for the Byzantine Agreement protocol
\item The randomness entropy $Q^r$ for selecting the next round leader and the next verifier set
\item Player replaceability in each step of the Byzantine Agreement protocol
\end{itemize}

As mentioned in the preceding paragraph, one of the major innovations of the ALGORAND is the use of an efficient 
BA protocol. Though it is a very efficient BA protocol, it incurs the major computational cost in ALGORAND implementation.
In this paper, we show that there is no need for ALGORAND to use BA protocols. A simple majority vote will achieve
the same goal as the BA protocol under the same assumption. 
There are several assumptions for the ALGORAND to work correctly. Among these assumptions, the most important ones 
are: more than $2/3$ of the users (or more than $2/3$ of the total money in permissionless environments) 
are honest and an honest user will delete her ephemeral private keys after each usage.
The authors of ALGORAND proved that under these two assumptions the probability for the block chain to fork 
is less than $1/10^9$ (it is claimed as $1/10^{12}$ or $1/10^{18}$ in the technical report \cite{chen2016algorand} though). 
In this paper, we will show that under the 
first assumption, the ALGORAND block chain may fork easily and we will show that the second assumption is not true for 
bribery attacks. For example, a malicious user can use bribery attacks to fork the block chain easily.

The structure of the paper is as follows. Section \ref{cryptsec} briefly describes the major cryptographic primitives 
for ALGORAND. Section \ref{algorandsec} briefly reviews the ALGORAND protocol. Section \ref{honemassumsec}
shows that honest majority users (or money) is not a good assumption for 
ALGORAND since one can fork the chain easily under this assumption. Section \ref{mafindhossec}
shows that the assumption that the majority individual users are honest is not realistic for ALGORAND.
Section \ref{byzalgosec} shows that Byzantine Agreement is not necessary for ALGORAND.

\section{Cryptographic primitives}
\label{cryptsec}
ALGORAND uses a digital signature scheme $SIG(\cdot)$ that satisfies the uniqueness property. 
Informally speaking, a signature scheme has the unique signing property if  it is computationally
infeasible to find a public key $pk$, a message $m$ and two values $s\not=s'$ such that 
$$VSIG(pk, m,s)=VSIG(pk, m, s')=1$$
That is, both $s$ and $s'$ could be verified as valid digital signatures on a single message $m$ using the public key $pk$.
Uniqueness signature schemes are used for permanent transactions on the block chain. 

In addition to the uniqueness signature scheme, identity based cryptographic schemes are used for 
message authentication during the block generation process and the private keys for identity 
based schemes are kept ephemeral. Specifically, when a user $i$ 
joins ALGORAND, the user  $i$ generates a public master key  $PMK_i$ and a corresponding
secret master key $SMK_i$. The user $i$ uses her long term uniqueness signature scheme key to digitally sign the 
public master key $PMK_i$. For each potential round-step pair $(i, r,s)$, the user $i$ computes
the identity based private key $sk_i^{r,s}$ for the public identity key $pk_i^{r,s}=(i, r,s)$. This private-public key
pair $(sk_i^{r,s},pk_i^{r,s})$ will be used for messages authentication during the step $s$ of round $r$. After these identity based
private keys $sk_i^{r,s}$ are generated, an {\em honest} user $i$ SHOULD destroy the secret master key $SMK_i$.

\section{A brief review of the ALGORAND protocol}
\label{algorandsec}
The initial status of the block chain is 
$$S^0=\left\{(1, a_1), \cdots, (j,a_j)\right\}$$
where $1, 2, \cdots, j$ are a list of initial users and $a_1, \cdots, a_j$ are their respective initial 
amounts of money units. We assume that each user $i$ is identified by its public key $pk_i$.
That is, for the users $1, 2, \cdots, j$, their corresponding public keys are $pk_1, \cdots, pk_j$.
A valid payment from a user $i$ to a user $i'$ is in the format  of 
$$SIG_{pk_i}(i, i', a')$$
where the user $i$ currently has $a\ge a'$ money units, $i'$ is an existing or a newly created user,
and $pk_i$ is the public key of user $i$. The impact of this payment is that the amount of money units for user
$i$ is decreased by $a'$ and the amount of money units for user $i'$ is increased by $a'$. 

In an idealized magic ledger system, all payments are valid and the list $L$ of sets of payments 
are posted in a tamper-proof box in the sky which is visible to all participants
$$L=PAY^0, PAY^1, PAY^2, \cdots,$$
ALGORAND block chain is organized in a series of rounds $r=0,1, 2, 3, \cdots$. Similar to the initial status,
the system status for round $r>0$ is a list of users and their corresponding money units
$$S^r=\left\{(1, a_1^{(r)}), (2,a_2^{(r)}), (3,a_3^{(r)}), \cdots\right\}$$
In a round $r$, the system status transitions from $S^r$ to $S^{r+1}$ via the payment set $PAY^r$
$$PAY^r: S^r\rightarrow S^{r+1}.$$

In ALGORAND, the block chain is a list of blocks $B^0, B^1, \cdots, B^r$ where each $B^r$ consists of the following fields:
the block number $r$ itself, the set $PAY^r$ of payments for round $r$, a quantity $Q^r$ which is the entropy seed 
for round $r$ randomness, the hash of the previous block $H(B^{r-1})$, and a set $CERT^r$ of signatures certifying 
that the block $B^r$ is constructed appropriately
$$B^r=\left\{r, PAY^r, Q^r, H(B^{r-1}), CERT^r\right\}.$$
It should be noted that in the ALGORAND protocol \cite{chen2016algorand,gilad2017algorand}, 
the field $CERT^{r-1}$ of the previous block $B^{r-1}$  is not included in the hash 
$H(B^{r-1})$. The field $CERT^{r}$ is a list of signatures for the value $H\left(r, PAY^r, Q^r, H(B^{r-1})\right)$
from at least $2/3$ of the members of the selected verifier set $SV^r$ for round $r$.

In ALGORAND, it is assumed that all messages are timely delivered in the entire network. 
Specifically, ALGORAND assumes that, at the start of round $r$, all users should have learned 
the current block chain $B^0, B^1, \cdots, B^{r-1}$. From this chain, one can deduce the user sets $PK^0, PK^1, \cdots, PK^{r-1}$ 
of each round. A potential leader of round $r$ is a user $i$ satisfying the condition
$$0.H(SIG_i(r,1,Q^{r-1}))\le p$$
where $p$ is a pre-determined probability chosen in such a way that, with overwhelming probability, at least one potential leader
is honest.  Note that the underlying signature scheme $SIG_i(\cdot)$ satisfies the uniqueness
property which requires that, given a message $m$, it is computationally infeasible to find two different signatures 
on the message $m$. Thus it is guaranteed that a user $i$ cannot increase his probability to be a leader by trying 
different signatures on the value $(r,1,Q^{r-1})$. Note that the user $i$ is the only person in the system that can determine
whether she is a potential leader since she is the only person that could compute the credential
$\sigma^r_i=SIG_i(r,1,Q^{r-1})$. However, the user $i$ can prove to others that she is a potential leader by releasing 
the credential $\sigma^r_i$. The leader $l^r$ is defined to be the user whose hashed credential is the smallest.
That is, $0.H(\sigma^r_{l^r})\le 0.H(\sigma^r_i)$ for all potential leaders $i$. Furthermore,
we also require that a user $i$ can serve as the leader in round $r$ only if she has joined the block chain 
$k$ blocks before where $k$ is a system-wide pre-determined parameter. That is, user $i$ can serve 
in round $r$ only if $i\in PK^{r-k}$. ALGORAND recommends $k=40$ in \cite{chen2016algorand}.

At the start of round $r$, each potential leader $i$ of round $r$ collects the maximal payment set $PAY^r_i$  
of round $r$ that have been propagated to her.
Then she computes the candidate block without the certificate $CERT^r$
$$B^r_i=\left\{r, PAY_i^r, SIG_i(Q^{r-1}), H(B^{r-1})\right\}$$
Next the user $i$ uses her identity based ephemeral private key $sk_i^{r,1}$ corresponding to the identity
public key $(i,r,1)$ to generate the following message
$$m_i^{r,1}=\left(B^r_i, ESIGN_{i,r,1}(H(B^r_i)), \sigma_i^r\right).$$
The user $i$ then destroys her ephemeral private key $sk_i^{r,1}$ and propagates the message $m_i^{r,1}$ to the entire network.

Since there could be several potential leaders during round $r$, each user could receive several candidate 
block messages $m_i^{r,1}$ from the step 1 of round $r$. Thus we need to select a set of verifiers to carry out 
the Byzantine Agreement protocol to determine the actual leader $l^r$ and the corresponding block 
$B^r_{l_r}$ proposed by this leader. Specifically, each step $s>1$ of round $r$ is executed by a set $SV^{r,s}$
of selected verifiers. A user $i\in PK^{r-k}$ belongs to the verifier set $SV^{r,s}$ if 
$$0.H(SIG_i(r,s,Q^{r-1}))\le p'$$
where $p'$ is a pre-determined probability such that the verifier set $SV^{r,s}$  satisfies certain conditions required by 
the ALGORAND. For example, the authors in  \cite{chen2016algorand} recommended choosing $p'$ in such a way that
the size of $SV^{r,s}$ is approximately 1500.

In the step 2 of round $r$, each verifier $i$ in $SV^{r,2}$ determines that the user $l$ is the round leader 
if $H(\sigma^{r,1}_l)\le H(\sigma^{r,1}_j)$
for all credentials $\sigma^{r,1}_j$ contained in the messages $m_j^{r,1}$ that she has received. After verifying 
the validity of the message $m_l^{r,1}=\left(B^r_l, ESIG_{l,r,1}(H(B^r_l)), \sigma_l^r\right)$, the verifier $i$ 
sets her initial value as $v'_i=H(B^r_l)$. The verifier $i$ uses her ephemeral identity based private key $sk_i^{r,2}$ to 
compute the message 
$$m_i^{r,2}=\left(ESIG_{i,r,2}(v'_i), \sigma_i^{r,2}\right),$$
destroys the ephemeral identity based private key $sk_i^{r,2}$, and propagates the message $m_i^{r,2}$ to the entire network.

From step $s=3$ to step $s=m+3$ (the authors in  \cite{chen2016algorand} recommended the value $m=180$),
the users in the verifier sets $SV^{r,s}$ execute the Graded Consensus Byzantine Agreement protocol to reach 
an agreement on the value $v'_i=H(B^r_l)$. From  $v'_i$, one can then determine the leader $l$ and 
the corresponding candidate block $B_r^l$ proposed by the leader $l$.
If a verifier $i$ determines that she has reached the agreement at step $s$ of round $r$, she would certify the block $B_r^l$ 
by generating a message 
\begin{equation}
\label{certmsg}
m_i^{r,s}=\left(ESIG_{i,r,s}(b_i),ESIG_{i,r,s}(H(B_r^l)), \sigma_i^{r,s}\right)
\end{equation}
where $b_i=0$ if the candidate block $B_r^l$ is not an empty block and $b_i=1$ otherwise.

The next block $B^r$ is finalized if the participants could collect at least $t_H$ valid 
certificates $m_i^{r,s}$ of format (\ref{certmsg}) where $t_H$ is a pre-determined system parameter.
Then the user attaches at least $t_H$ certificates $m_i^{r,s}$ of format (\ref{certmsg}) as $CERT^r$ to the block $B^r$.
In the case that $B^r$ is an empty block (i.e., $PAY^r=\emptyset$), then the value $Q^r$ is defined as $Q^r=H(Q^{r-1}, r)$.

\section{Honest majority assumption: Wrong!}
\label{honemassumsec}
In permissioned ALGORAND environments, it is assumed that at least $2/3$ of the users are honest and 
in permissionless ALGORAND environments, it is assumed that at least $2/3$ of the money units are honest.
Under these assumptions, it was ``proved'' that the probability for the ALGORAND block chain to fork is at most $1/10^9$.

In the following, we show that these assumptions will not guarantee the fork-free property for the ALGORAND 
block chain. Indeed, it is very efficient for certain adversaries who control at most $1/3$ of the users (or money units) 
to fork the block chain.

Assume that the current block chain is 
$$B^0, B^1, \cdots, B^{r_1}, \cdots, B^r$$
and the corresponding user sets are
$$PK^0, PK^1, \cdots, PK^{r_1}, \cdots, PK^r$$
Furthermore, assume that $3|PK^{r_1}|<|PK^r|$.

The adversary ${\cal A}$ chooses to corrupt all users in  $PK^{r_1}$. Since $\frac{|PK^{r_1}|}{|PK^r|}<\frac{1}{3}$, 
${\cal A}$  is a valid adversary in the ALGORAND adversary model. Since ${\cal A}$
controls all users in the block $B^{r_1}$, she can begin to fork the chain from $B^{r_1}$ and construct a new chain 
$$B^0, B^1, \cdots, B^{r_1}, \bar{B}^{r_1+1},\cdots, \bar{B}^r,  \bar{B}^{r+1}$$
where $\overline{PK^j}=PK^{r_1}$ for $r_1<j\le r$, $\overline{PK^{r+1}}=PK^{r}$, 
the payment set $\overline{PAY^{j}}$ for $r_1<j\le r$ consists of some simple transactions among the users
within $PK^{r_1}$, and the payment set 
$$\overline{PAY^{r+1}}=\left\{SIG_{pk_{i_1}}\left(i_1, i_1',a_1\right), \cdots, SIG_{pk_{i_t}}\left(i_t, i_t',a_t\right)\right\}$$
where $\{i_1, \cdots, i_t\}\subseteq PK^{r_1}$, $PK^r\setminus PK^{r_1}=\{i'_1, \cdots, i'_t\}$, 
and $a_1, \cdots, a_t$ are tiny amounts of money units. 
It should be noted that this forked chain could be generated very efficiently since the 
adversary ${\cal A}$ does not need to collect other transactions and does not need to carry out Byzantine Agreement.
All she needs to do is to find sufficient number of users in $PK^{r_1}$ under her control to certify the 
blocks $\bar{B}^{r_1+1},\cdots, \bar{B}^{r+1}$ which could be done very efficiently.

The forked chain $B^0, B^1, \cdots, B^{r_1}, \bar{B}^{r_1+1},\cdots, \bar{B}^r,  \bar{B}^{r+1}$ is longer than the original chain
$B^0, B^1, \cdots, B^{r_1}, \cdots, B^r$. Thus the forked chain should be adopted as the legal chain and the fork is successful.

The example in the preceding paragraphs shows that it is not realistic to assume  that the majority 
users (or majority money units) are honest. Thus it is important to investigate other realistic assumptions 
for making block chains fork-free. 
Indeed, our attack shows that for proof-of-stake based block chains, the users within the genesis block 
can collectively re-build the entire block chain at any time. Thus  we recommend block chain techniques 
that use  a combination 
of proof-of-work (or proof-of-something that is hard to be efficiently done) and proof-of-stake.
Ethereum is a good example of block chains that will adopt a combination of proof-of-stake and proof-of-work.

\section{Majority of individual users are honest: Wrong!}
\label{mafindhossec}

In ALGORAND, it is assumed that majority users (or majority money units) are honest. In particular, it is assumed 
that all honest users will destroy the ephemeral private keys after these keys have served their purpose. This assumption 
is not realistic in practice. In the seminal work of Bitcoin, it is assumed that each individual user could be 
malicious though an adversary may not be able to coordinate more than 50\% of the computing resources 
within the entire network. In a distributed environment (in particular, in a permissionless block chain network),
it is not realistic to assume that an individual users is {\em honest}. In a public network, we have to 
assume that every participant is ``curiously malicious''.  In particular, 
if there is an incentive for an individual participant to take a certain action, why would she refuse?
For most users (if not all) in the block chain network, they would take advantages of their leadership and 
verifier roles if they were selected to serve.

In ALGORAND, secret cryptographic sortition techniques are used to select potential leaders and verifiers set
in a secret way. That is, only the selected users learn the fact that they are selected. 
The adversary does not know which users to corrupt since she does not know who would be the leader
and who would be the verifiers. However, the adversary can provide sufficient incentives to ask 
selected leaders and verifiers to publish their roles before they serve on their roles. In this way,
the adversary could identify the target users to corrupt before the protocol continues. Thus 
the leaders and the verifiers would produce messages in favor of the adversary.

In ALGORAND, it is also assumed that honest users would destroy their ephemeral keys after these keys 
have been used to authenticate corresponding messages. By requiring this, the adversary will not 
be able to ask the leader/verifiers to fork the chain from the previous block since the 
leader/verifiers do not have the ephemeral keys to certify the block any more.
In particular, it is mentioned in \cite{chen2016algorand} that 
\begin{quote}
``Roughly, once $B^r$ has been generated, the Adversary has learned who the verifiers of each step of round $r$ are. 
Thus, he could therefore corrupt all of them and oblige them to certify a new block  $\tilde{B^r}$......We 
do so by means of a new rule. Essentially, the members of the verifier set $SV^{r,s}$ of a step $s$ of 
round $r$ use ephemeral public keys $pk^{r,s}$ to digitally sign their messages. These keys are 
single-use-only and their corresponding secret keys $sk_i^{r,s}$ are destroyed once used. This way, if a verifier is
corrupted later on, the Adversary cannot force him to sign anything else he did not originally sign.''
\end{quote}
Obviously there is no incentive for an individual user
to destroy her ephemeral private keys after the keys have been used.
If she does not destroy her ephemeral private keys, she may be able to sell these keys
to the adversary later. Thus there is an incentive for her not to destroy these keys.

In a summary, the assumption that the majority of ``individual users'' are honest is not true in practice. 
Everyone in the block chain network would like to maximize her benefit (why not?). 
The selected leaders and verifiers would accept bribery if it is attractive enough. 
There is no incentive for an individual user to keep her leadership role private. She would 
rather broadcast her roles before she serves on these roles so that she could accept bribery.
She would not destroy her private keys so that she could sell them later.
In other words, it is reasonable to assume that the majority of the users in the entire network 
is not coordinated by a single adversary (cf. Bitcoin assumption) though it is not reasonable 
to assume that the majority of individual users in the network are not individually malicious.

\section{Byzantine Agreement is not necessary in ALGORAND}
\label{byzalgosec}
In ALGORAND, Byzantine Agreement (BA) is used to reach consensus on the next block $B^r$ and to avoid forks.
Specifically, after the first step of round $r$, selected verifiers in the set $SV^{r,2}$ received multiple 
proposals for the next block $B^r$ from multiple potential leaders. Every verifier $i$ in $SV^{r,2}$
selects the qualified leader  $l_i$ from the pool of potential leaders by comparing their credential hash outputs,
and extracts the candidate block $B_i^r$. Let $v_i'=H(B_i^r, SIG_{pk_{l_i}}(B_i^r))$.
At this stage, different verifiers in $SV^{r,2}$ may have selected different proposals for the block $B^r$
and holds different values $v_i'$. Thus they need to use a Byzantine Agreement protocol to reach a consensus on $v_i'$
(from which they can reach consensus on $B^r$).
The BA protocol in ALGORAND is based on the Graded Consensus (GC) and proceeds as follows
where we assume that, at each step, more than $2/3$ participants are honest. 
\begin{enumerate}[label=(\Alph*)]
\item\label{bsteepA} Each user $i\in SV^{r,2}$ propagates her authenticated value $v_i'$ to the entire network
\item\label{bbstep2} Each user $i\in SV^{r,3}$ propagates the authenticated string $x$ to 
the entire network if and only if she received the string $x$ from  more than $\frac{2|SV^{r,2}|}{3}$ users
\item\label{mstep3} Each user $i\in SV^{r,4}$ calculates $(v_i,g_i)$ as follows:
\begin{itemize}
\item If she received the string $x$ from more than $\frac{2|SV^{r,3}|}{3}$ users during Step \ref{bbstep2}, then $v_i=x$ and $g_i=2$
\item\label{bstep5}  If she received the string $x$ from more than $\frac{|SV^{r,3}|}{3}$ users during Step \ref{bbstep2}, 
then $v_i=x$ and $g_i=1$
\item Otherwise, $v_i=\emptyset$ and $g_i=0$
\end{itemize}
\item All users $i\in SV^{r,s}$ with $s\ge 4$ execute the Binary Byzantine Agreement (BBA) Protocol with the following input values:
Each user $i\in SV^{r,4}$ sets her initial input value for BBA as $0$ if $g_i=2$ and $1$ if $g_i<2$.
\item After the BBA protocol, each participant $i$ outputs $v_i$ if the output from BBA is $0$. Otherwise, outputs $\emptyset$.
\end{enumerate}

From the above BA protocol, it is clear that the honest users agree on a non-empty block $B^r$ 
only if the BBA protocol output is $0$ for all honest participants.
This happens only if at least one of the honest participants in Step \ref{bstep5} holds a value $(v_i, 2)$. Again, this happens 
only if the first sub-step in Step \ref{mstep3} happens for at least one honest participant. 
This means that at least $2/3$ of users $i\in SV^{r,3}$ in the Step \ref{bbstep2} has signed the block candidate $B^r$.
This again means that more than $\frac{2|SV^{r,2}|}{3}$ users from $SV^{r,2}$ signed $B^r$ in Step \ref{bsteepA}.
By these facts, 
we do not need to carry out the BA protocol for ALGORAND to get consensus on $B^r$. Instead, the consensus 
on $B^r$ could be reached very efficiently as follows.
\begin{enumerate}[label=(\alph*)]
\item Step 2 of round $r$: each verifier $i$ in $SV^{r,2}$ authenticates and propagates 
her candidate block $B^r_i$ to the entire network
\item\label{simstep2} All users check whether she has received more than $\frac{2|SV^{r,2}|}{3}$ signatures for some candidate block 
$B^r$. If there exist more than $\frac{2|SV^{r,2}|}{3}$ signatures for a proposed block 
$B^r$, then marks $B^r$ as the final round $r$ block.
\end{enumerate}

It should be noted that the above simplified two-step protocol will achieve the same goals as the Byzantine 
Agreement protocol in ALGORAND. First note that if an honest user determines in Step \ref{simstep2} that a block 
$B^r$ has been certified by more than $\frac{2|SV^{r,2}|}{3}$ users in $SV^{r,2}$,  then no user
in Step \ref{simstep2} will receive more than $\frac{2|SV^{r,2}|}{3}$ signatures for any other candidate blocks
since there are at most $\frac{|SV^{r,2}|}{3}$ malicious users in $SV^{r,2}$. This proves that no fork for the block chain
will happen (this is the major goal  for the ALGORAND Byzantine Agreement protocol). Furthermore,
our analysis in the preceding paragraphs show that if ALGORAND Byzantine Agreement protocol would 
agree on a block $B^r$, then the above simplified two-step protocol would also agree on the same block  $B^r$.

\section{Other related Proof of Stake protocols}
In the Sleepy Consensus Model (Pass and Shi \cite{pass2017sleepy}), a user $i$ gets the priority to produce the next 
block if $H(i,t)<p$ where $i$ is the user identity, $t$ is the current time stamp, and $p<1$
is a pre-determined probability. This Nakamoto-style protocol requires weakly synchronized clocks
and forks are created frequently. Furthermore, for malicious adversaries, our attacks in Sections  \ref{honemassumsec}
and  \ref{mafindhossec} work against this Sleepy Consensus Model protocol also. 
It should be noted that, in the adaptive security model of \cite{pass2017sleepy}, 
a user $i$ gets the priority to produce the next block if ${\tt PRF}_k(t)\oplus {\tt PRF}_{k_0}(i,t)<p$
where $k_0$ is a random seed included in the common reference string and $k$ is a secret key that the user $i$ 
has committed to the public key infrastructure.

Kiayias et al \cite{kiayias2017ouroboros} claims that  their 
Ouroboros protocol is the first proof-of-stake block chain protocol with a provable security. It is straightforward to check
that our attacks in Section \ref{honemassumsec} works against the Ouroboros protocol  also.

\bibliographystyle{plain}


\end{document}